\documentclass[preprint,aps,pra,showkeys,showpacs]{revtex4}
\usepackage{tabularx}
\usepackage{dcolumn}
\usepackage{graphics}
\usepackage{bm}
\usepackage[dvips]{graphicx}
\usepackage{dcolumn}
\usepackage{tabularx}
\usepackage{amsmath}
\usepackage{amssymb}
\usepackage{xcolor}

\begin{document}

\title{$\bar{D}^{0}D^{0*}$ $(D^{0}\bar{D}^{0*})$ System in QCD-Improved Many Body Potential}
\thanks{BM and FA acknowledge the support of PU research grant no.
D/34/Est.1 Sr. 108 and 109 respectively. SG is thankful to the
Higher Education Commission (HEC) of Pakistan for its financial
support through Grant No. 17-5-4(Ps3-128) HEC/Sch/2006.}
\author{M. Imran Jamil}
\affiliation{University of Management and Technology, Lahore,
Pakistan.}
\author{Bilal Masud}\affiliation{Centre For High Energy Physics, Punjab University,
Lahore(54590), Pakistan.}
\author{Faisal Akram}\affiliation{Centre For High Energy Physics, Punjab University,
Lahore(54590), Pakistan.}
\author{S. M. Sohail Gilani}
\thanks{email:msgilani2005@gmail.com}\affiliation{Centre For High Energy Physics, Punjab University, Lahore(54590), Pakistan.}
\thanks{bilalmasud.chep@pu.edu.pk, faisal.chep@pu.edu.pk, msgilani@hotmail.com}
\date{\today}
\begin{abstract}
For a system of current interest (composed of charm, anticharm quarks and a pair of light ones),
we show trends in phenomenological implications of QCD-based improvements to a simple quark model treatment.
We employ resonating group method to render this difficult four-body problem manageable.
 We use a quadratic confinement so as to be able to improve beyond the Born approximation.
 We report the position of the pole corresponding to
$\bar{D}^{0}D^{0*}$ molecule for the best fit of a model parameter to the relevant QCD simulations.
We point out the interesting possibility that the pole can be shifted to
 $3872$ MeV by introducing another parameter $I_{0}$ that changes the strength of the
interaction in this one component of $X(3872)$. The revised value of this second parameter can guide future trends in modeling of the full
exotic meson $X(3872)$.
We also report the changes with $I_{0}$ in the $S$-wave spin averaged cross sections for
$\bar{D}^{0}D^{0*}\longrightarrow\omega J/\psi$ and $\bar{D}^{0}D^{0*}\longrightarrow\rho J/\psi$.
These cross sections are important regarding the study of QGP (quark gluon plasma).
\end{abstract}


 \maketitle

\section{Introduction}
Considering difficulties in solving quantum chromodynamcis (QCD) for
the relevant energies, hadron phenomenology and hadron-hadron
scattering is studied mostly through models or effective Lagrangian
densities. But as far as possible continuum hadronic models should
agree to lattice simulations of QCD and give phenomenological
implications having a good comparison with the corresponding hard
experimental results. For multiquark systems, a common approach
having a fairly good phenomenological record, is the sum of
pair-wise interaction model \cite{J. Weinstein, diagrammatic
approach, Gui-Jun Ding Wei Huang Jia-Feng Liu Mu-Lin Yan,Eric S
Swanson,Y. Cui X-L. Chen, T. Barnes N. Black,Wong Swanson
Barnes,Wong Swanson 01,Bicudo,Swanson report 06,Wang Huang,Hiyama
prog th physics,Hiyama physics letter B}. The need for improvement
in it is indicated even phenomenologically by noting that this model
predicts color van der Waals interaction of the inverse-power type
between separated hadrons and this has no experimental evidence. At
the quark level, good lattice-based improvements \cite{B. Masud, A.
M. Green and P. Pennanen, P. Pennanen,green2} to this sum of
two-body potential model are available which modify it at large
distances. These improvements introduced a space dependent form
factor $f$ (appearing in eqs. \eqref{e3.7a}, \eqref{e3.7b} and
\eqref{e15} below) in off-diagonal elements in the overlap,
potential and kinetic energy matrices of the model. The additional
parameter in $f$ minimizes difference between the two quark two
antiquark binding in the improved model to the binding resulting
from relevant lattice-generated QCD simulations by UKQCD
\cite{A.M.Green and C.Michael and M.E.Sainio, A.M.Green and J.
Lukkarinen and P. Pennanen and C.Michael and S.Furui,A.M.Green and
J.Lukkarinen and P. Pennanen and C. Michael, Petrus Pennanen}. The
exponential form of $f$ keeps the model agreeing to the pair-wise
interaction model in the small distance limit while getting a fairly
good agreement to the QCD simulations and solving the van der Waals
problem.

It is necessary to find testable implications of these improvements at the meson level in form of multiquark energies (binding)
and meson-meson cross-sections. Without these improvements, the $\bar{D}^{0}D^{0*}$ and
its coupling to $\omega J/\psi$ or $\rho J/\psi$ has been studied \cite{Wong Swanson Barnes,Wong Swanson 01,Cheuk-Yin Wong}.
Ref.  \cite{Wong Swanson Barnes,Wong Swanson 01} report the resulting  $\rho J/\psi$ to $\bar{D}^{0}D^{0*}$ cross sections, along with many others.
Ref. \cite{Cheuk-Yin Wong} reports meson-meson potential and eigenvalues for $D\bar{D}^*$ and $B\bar{B}^*$ four-quark states
and find molecular states in the resulting combinations. We are now calculating revised implications for the $\bar{D}^{0}D^{0*}$ system.
These implications address some experimental issues of wide interest, for example understanding exotic mesons \cite{Godfrey,Close Vijande 04,Shi-Lin Zhu}.
An important such state is the meson $X(3872)$
which is now generally considered
\cite{y Dong and a Faessler,Ortega and Segovia PRD, Bugg, Eitchen Lane, Voloshin 05,Lee I W and Faessler A and Gutsche T and Lyubovitskij V E,Kalashnikova Yu S and Nefediev A V}
as a mixture of $\bar{D}^{0}D^{0*}$, $D^{+}D^{-*}$ and $c\bar{c}$.
Any effort to understand it, thus, should understand quantities depending upon its components. A direct lattice
QCD study of it would have to calculate many Wilson loops before
arriving at any conclusion. A more manageable route could be to make
separate models of its components, find out their consequences and
then combine the models to understand $X(3872)$. Our work is the first step in this scheme; we take up $\bar{D}^{0}D^{0*}$
system whose flavor content has an overlap with both isovector $\rho J/\psi$ and isoscalar $\omega J/\psi$ and we study its coupling to both channels.

Ref. \cite{Swanson 04} addresses the possibility that $X(3872)$ is a molecular bound state of neutral charm mesons and refs.
\cite{Braaten a5, Braaten a7, P. Colangelo and F. De Fazio and S. Nicotri, Masayasu Harada and Yong-Liang Ma, T. Mehen and R. Springer}
assume so. Ref. \cite{Eric S Swanson} says that $\bar{D}^{0}D^{0*}$ to  $\omega J/\psi$ (and $\rho J/\psi$) interaction is needed
to understand models of $X(3872)$. $\bar{D}^{0}D^{0*}\longrightarrow\omega (\rho) J/\psi$ scattering is needed to understand the final state interaction
in the $X(3872)$ decaying to  $J/\psi\rho$ or $J/\psi\omega$ through the intermediate $\bar{D}^{0}D^{0*}$.  Refs. \cite{Liu Zang Zhu 06,Meng Chao 07}
describe the role of this final state interaction through the effective lagrangian approach. We present results that may have implications for these final
state interactions while being closer to QCD in giving a quark level description.  Refs. \cite{Braaten
and Kusunoki and Nussinov, Braaten and Kusunoki} use the sub-process
$\bar{D}^{0}D^{0*}\longrightarrow\bar{D}^{0}D^{0*}$ for
the final state interaction in net $B\longrightarrow\bar{D}^{0}D^{0*}K$
process. Our comments also apply to this channel and we have shown below our results for
$\bar{D}^{0}D^{0*}\longrightarrow\bar{D}^{0}D^{0*}$
scattering as well. In a recent paper, Braaten and Kang \cite{Braaten and Kang} say that ``in case of $1^{++}$ quantum numbers of $X(3872)$,
effects of scattering between  $\omega J/\psi$ and charm meson pairs could be significant."
Moreover, $\bar{D}^{0}D^{0*}\longrightarrow\omega (\rho) J/\psi$ scattering is needed for studying the effect of final state interaction between the
comovers in relativistic heavy ion collision experiments  \cite{Barnes in EPJA}.

For the $\bar{D}^{0}D^{0*}$ system, another improvement beyond the quark-antiquark pair-wise interaction implemented  is  ref. \cite{Eric S Swanson}.
This adds a point-wise meson interaction to the coupling resulting from one gluon exchange and calculates the resulting $\bar{D}^{0}D^{0*}$ to $\omega J/\psi$
scattering amplitudes. We, in this paper, present $\bar{D}^{0}D^{0*}$ to  $\omega J/\psi$ and $\rho J/\psi$ cross-sections along with an analysis of
$\bar{D}^{0}D^{0*}$ binding resulting from the $f$ model \cite{A. M. Green and P. Pennanen, P. Pennanen,green2} that better fits the available QCD
simulations than the one gluon exchange model. In a previous work \cite{M. Imran Jamil and Bilal Masud}, we used Born approximation to calculate the
meson-level consequences of the most developed geometrical form of the $f$ factor. In the present paper, we use a resonating group formalism to avoid
the Born approximation used in refs. \cite{M. Imran Jamil and Bilal Masud, T. Barnes E.S. Swanson J. Weinstein, T. Barnes E.S. Swanson 94, T.Barnes NuovoCim}
for meson-meson scattering and thus report results can be compared
with Born approximation \cite{Gilani imran bilal faisal}. This is essential to be able judge how good is this approximation.
To get analytic expressions for the resulting scattering amplitudes,  now we use a quadratic confinement and a simpler form of the $f$ factor.
We incorporate the spin and flavour dependence. A similar realistic meson-meson treatment for lighter quarks was published earlier \cite{masud}.
We now address a system ($\bar{D}^{0}D^{0*}$) of current interest and give a much more thorough analysis of the meson-meson binding. Moreover,
we include the meson-meson cross-sections that are not in \cite{masud} at all.

These cross sections can be useful in the experimental studies of quark-gluon plasma (QGP) in relativistic heavy ion collisions. One of the promising
signature of QGP in heavy ion collision experiments is the suppression of $J/\Psi$ caused by color Debye screening. However the observed suppression may
be affected by the interaction of $J/\Psi$ with the comoving Hadrons mainly $\pi$ and $\rho$ Mesons after the hadronization of QGP. The effect of the
interaction with the comovers can be significant as the density of these mesons is very high. Thus an estimate of these cross sections can help in identifying
any contribution of QGP in observed production rate of $J/\Psi$ in heavy ion collision experiments.

This paper is organized as follows. In Section \ref{sec2} we have
specified our $q^2\bar{q}^2$ Hamiltonian
and written the
spin and flavor wave functions and the form of the position wave function
of our system. The section ends with the integral equations for the unknown
position factors of our total wave function, as in a resonating group formalism.
In Section \ref{solveintegraleqs}, we solve our integral equations for the
amplitudes of transition between two channels of our multiquark system.  In Section \ref{fitparameters} we report the best fit values of the parameters used
in our formalism
along with describing how they are fixed. In Section \ref{sec5}, we present our results
for the scattering cross-sections and bindings
and give conclusion.
\section{The Hamiltonian Matrix and the
Wave Functions}\label{sec2}
We use the adiabatic approximation to first define the potential for fixed positions of two quarks and two antiquarks.  \noindent The model we use
(of ref. \cite{green2}, with position dependence as that of the model $I_a$ in ref. \cite{P. Pennanen}) improve the kinetic, potential and overlap matrices
in the color basis
\begin{equation}
|1\rangle_{c}=|1_{1\bar{3}}1_{2\bar{4}}\rangle_{c}, \\
\hspace{.25in} |2\rangle_{c}=|1_{1\bar{4}}1_{2\bar{3}}\rangle_{c}.
\label{basis}
\end{equation}
They fit to the lattice simulations a parameter $k_{f}$ introduced in the off-diagonal position dependent elements of these matrices, while keeping
the small distance limit of the model agreeing to the pair-wise model. To avoid Born approximation, we had to use the simplest form
 \begin{equation}f=\text{exp}(-b_{s} k_{f}\sum_{i<j} r_{ij}^{2}).\label{e7}\end{equation}
in the off-diagonal elements that is used in otherwise more developed model version in ref. \cite{green2}.

In the next step of the adiabatic approximation, we calculate quark position wave functions. For this, we start by writing our total state vector as a
sum over $k$ of product of the gluonic states $|k\rangle_{g}$, known spin and flavor states and the corresponding quark position wave function
$\Psi^k(\textbf{r}_1,\textbf{r}_2,\textbf{r}_{\bar{3}},\textbf{r}_{\bar{4}})$. $|k\rangle_{g}$ is defined as
QCD eigenstate that approaches the corresponding colour state $|k\rangle_{c}$  in the small distance limit. The position dependence of the overlaps and
potential energy matrices in the $\{|k\rangle_{g}\}$ basis are taken from the above mentioned refs. \cite{P. Pennanen,green2}. For the kinetic energy
matrices we use the non-relativistic prescriptions used in ref. \cite{B. Masud}; there it is justified through effective hadron Hamiltonian \cite{thomas}
in (space-)lattice QCD. To these we add (after multiplying the appropriate identity matrices) the sum of the corresponding constituent quark masses
$m_{i}$ $(i=1,2,\bar{3},\bar{4})$, fixed \cite{ES Swanson} to meson spectroscopy, to get the total meson-meson Hamiltonian matrix; this semi-relativistic
prescription is already used in refs. \cite{J. Weinstein, diagrammatic approach, B. Masud, masud}. The resulting matrices are improvements to the matrices
in basis of eq. (\ref{basis}) of the Hamiltonian appearing in ref. \cite{J. Weinstein}, i.e.
\begin{equation}
\hat{H}=\sum_{i=1}^{\overline{4}}\Big[m_{i}+\frac{\hat{P}_{i}^{2}}{2m_{i}}\Big]+
\sum_{i<j}v(\textbf{r}_{ij})\mathbf{F}_{i}.\mathbf{F}_{j}.\label{e9}
\end{equation}
$\mathbf{F}_{i}$ is the set of color matrices (of $SU(3)_c$) for the ith particle. $\textbf{F}$ has $8$ components
$F_{a}=\frac{\lambda_{a}}{2}$ for a quark and for
an anti quark $F_a=-\frac{\lambda_{a}^*}{2}$, $a=1,2,3,...,8$. For using our analytic formalism beyond the Born approximation we employed a simple
harmonic potential already used in refs. \cite{J. Weinstein, B. Masud, masud}
\begin{equation}v(\textbf{r}_{ij})=v_{ij}=C
r_{ij}^{2}+\bar{C} \text{ with }
i,j=1,2,\bar{3},\bar{4},\label{e16}\end{equation}
rather than more sophisticated forms of refs.~\cite{Godfrey Isgur 85,Cheuk-Yin Wong, Li Chao}.
Our neglect of the hyperfine interaction is less serious in $\bar{D}^{0}D^{0*}\rightarrow\omega (\rho)J/\psi$ processes; ref.
\cite{Eric S Swanson} shows that this amplitude is dominated by the confinement interaction.

This specifies our formula of color interactions between different quarks. The explicit color dependent factor in it is $\mathbf{F}_{i}.\mathbf{F}_{j}$
and that is flavor independent in consistent with the color charge on a quark on any flavor being same. Its quadratic confining coefficient $C
r_{ij}^{2}+\bar{C}$ is to replace the more sophisticated forms of refs.~\cite{Godfrey Isgur 85,Cheuk-Yin Wong, Li Chao} in which the coefficient of the
confining term, the QCD \emph{string tension}, is everywhere taken to be flavor independent; the string tension models the energy density of the gluonic
field originating from color charges and color charges are same for each flavor. The confining term we use is the $C r_{ij}^{2}$ and its coefficient $C$
is accordingly taken to be flavor independent. This gluonic field energy density is calculated in the lattice QCD simulations of ref.~\cite{Fumiko Okiharu}
and this work advocates a flavor independent string tension. The constant term $\bar{C}$ is added to the flavor dependent sum of constituent quark masses in
 our actual formulas for meson masses, for example in eq. \eqref{e3.137} below.

As in the resonating group method, we factorize
$\Psi^k$ into known and unknown factors to utilize the well known SHO position wave functions $\xi_{k}(\textbf{y}_{k})$ and $\zeta_{k}(\textbf{z}_{k})$
within each quark antiquark subsystem
\begin{equation}|\Psi(\textbf{r}_1,\textbf{r}_2,\textbf{r}_{\bar{3}},\textbf{r}_{\bar{4}};g)\rangle=
\sum_{k=1}^2|k\rangle_{g}|k\rangle_{f}|k\rangle_{s}\Psi_{c}(\textbf{R}_{c})\chi_{k}(\textbf{R}_{k})\xi_{k}(\textbf{y}_{k})\zeta_{k}(\textbf{z}_{k}).\label{e11}\end{equation}

\begin{figure}[!h]
\includegraphics[scale=.90,angle=-0]{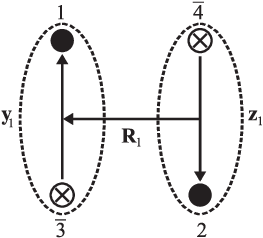}
\caption{Topology 1. } \label{fig1}
\end{figure}

\begin{figure}[!h]
\includegraphics[scale=.90,angle=-0]{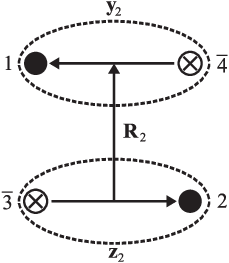}
\caption{Topology 2 } \label{fig2}
\end{figure}

Where $|k\rangle_{f}$ are the flavor states and  $|k\rangle_{s}$ are the spin states. Here  $\textbf{R}_{c}$ is the c.m. position vector. The inter-cluster vector
$\textbf{R}_{k}$ and
in-cluster vectors $\textbf{y}_{k}$ and
$\textbf{z}_{k}$ are shown in figs. \ref{fig1} and \ref{fig2}, which also define the topologies $k=1,2$. For example,
\begin{equation}\textbf{R}_{1}=\frac{(\textbf{r}_{1}+r \textbf{r}_{\overline{3}}-r
\textbf{r}_{2}-\textbf{r}_{\overline{4}})}{(1+r)}.\label{e4a}\end{equation}
Here $r=\frac{m_{c}}{m}$, with
 $m, m_{c}$ being the constituent mass of light (up or down) and charm quarks respectively.

The sizes $d_{k1}$ and $d_{k2}$ of the known quark antiquark clusters are also parameters of our model.  $d_{k1}$ is defined by
\begin{equation}\xi_{k}(\textbf{y}_{k})=\frac{1}{(2\pi
d_{k1}^{2})^{\frac{3}{4}}}\text{exp}\Big(\frac{-\textbf{y}_{k}^{2}}{4
d_{k1}^{2}}\Big).\label{Gaussian1} \end{equation}
$d_{k2}$ replaces $d_{k1}$ in $\zeta_{k}(\textbf{z}_{k})$.
 The unknown inter-cluster factor $\chi_{k}(\textbf{R}_{k})$ is our variational function found by solving integral eq. \eqref{e3.6} for it.
 To get this equation, we set the overlap of an arbitrary variation $|\delta \Psi\rangle$, in $|\Psi\rangle$ of eq. \eqref{e11},
 with $(\hat{H}-E_{c})|\Psi\rangle$ as zero and reading off the coefficients of the arbitrary variations $\chi_{k}(\textbf{R}_{k})$ with $k=1,2$. This gives
  \begin{equation}
\sum_{l=1}^2\int d^3\textbf{y}_{k}d^3\textbf{z}_{k} \hspace{.05 in} {}_{f}\langle
k|l\rangle_{f}\hspace{.05 in} {}_{s}\langle
k|l\rangle_{s} \hspace{.05 in} \xi_{k}(\textbf{y}_{k})
\zeta_{{k}}(\textbf{z}_{k})_{g}\langle k|
\hat{H}-E_{c}|\l\rangle_{g}\chi_{l}(\textbf{R}_{l})\xi_{l}(\textbf{y}_{l})
\zeta_{l}(\textbf{z}_{l})=0.\label{e3.6}
\end{equation}
The trivial integration over the c.m. position $\textbf{R}_{c}$
could be performed to give a finite result (implied in above
equation) using, say, a box normalization. It is to be noted that
our total meson-meson Hamiltonian is an identity operator in the
flavor and  spin basis because it differs from that in eq.
\eqref{e9} only through the position dependent $f$ and we are
neglecting the spin-spin hyperfine interaction.

We use
$_{g}\langle k|\l\rangle_{g}$, $_{g}\langle k| \Hat{V} |\l\rangle_{g}$ and
 $_{g}\langle k| \Hat{K} |\l\rangle_{g}$
of  refs. \cite{green2, B. Masud} to get $_{g}\langle k| \Hat{H}-E_c |\l\rangle_{g}$
 required in eq. (\ref{e3.6}).
These form the matrices:
\begin{equation}N\equiv \{ N_{kl}\}\equiv \{_{g}\langle k \mid l\rangle_{g}\}=\left(
            \begin{array}{cc}
              1 & \frac{1}{3} f \\
              \frac{1}{3} f  & 1 \\
            \end{array}
          \right),\label{e3.7a}
\end{equation}
\begin{equation}V\equiv \{ V_{kl}\}\equiv \{_{g}\langle k \mid \Hat{V} \mid l\rangle_{g}\}~~~~~~~~~~~~~~~~~~~~~~~~~~~~~~~~~~~~~~~~~~~~~~~~~~~~~~~~~~~~~~~~~~~~~~~~~~~~~~~~~~\nonumber\end{equation}
\begin{equation}=\left(
            \begin{array}{cc}
              -\frac{4}{3} (v_{1\overline{3}}+v_{2\overline{4}}) & \frac{4}{9} f (v_{12}+v_{\overline{3}\overline{4}}-v_{1\overline{3}}-v_{2\overline{4}}-v_{1\overline{4}}-v_{2\overline{3}})\\
             \frac{4}{9} f (v_{12}+v_{\overline{3}\overline{4}}-v_{1\overline{3}}-v_{2\overline{4}}-v_{1\overline{4}}-v_{2\overline{3}}) & -\frac{4}{3} (v_{1\overline{4}}+v_{2\overline{3}})
            \end{array}
          \right)\label{e3.7b}
  \end{equation}

\begin{equation}
K\equiv \{ K_{kl}\}\equiv
{}_{g}\langle k| \Hat{K}|\ l\rangle_{g}= N(f)_{k,l}^{\frac{1}{2}}
\Big(\sum_{i=1}^{\overline{4}}-\frac{\nabla_{i}^{2}}{2m}\Big)N(f)_{k,l}^{\frac{1}{2}}.\label{e15}
\end{equation}

For $\bar{D}^{0}D^{0*}$ (chosen as channel 1 with $k=1$), the total spin is 1. Angular momentum conservation
tells that in the quark exchanged channels
($\omega J/\psi$ and $\rho J/\psi$ corresponding to $k=2$ the total spin should be 1.
These spin states are denoted by
\begin{equation}|1\rangle_{s}=|P_{1\overline{3}}V_{2\overline{4}}\rangle\label{e3.1}\end{equation}
\begin{equation}|2\rangle_{s}=|V_{1\overline{4}}V_{2\overline{3}}\rangle\label{e3.2},\end{equation}
where $P$ represents a pseudo-scalar and $V$ represents a vector
meson. We utilized the rotational symmetry of our problem to write each of these \textit{S}=1  states
 as $\frac{1}{\sqrt{3}}\left(|1,1\rangle+|1,0\rangle+|1,-1\rangle\right)$
 \noindent with the second label as the $S_z$ quantum number. We then used the completeness of the meson and then quark spins, along with the required
 Clebsch-Gordan coefficients, to arrive at the following for ${}_{s}\langle
k|l\rangle_{s}$ in eq.(\ref{e3.6})
 \begin{equation} _{s}\langle 1|2\rangle_{s}= _{s}\langle 2|1\rangle_{s}= \frac{1}{\sqrt{2}}.\label{spin overlap} \end{equation}

\noindent The flavor content of our channel-1 is unique
\begin{equation}|1\rangle_{f}=|\bar{c}u\rangle|c\bar{u}\rangle.\label{e3.3}\end{equation}
For the second channel, it depends on our choice of mesons in it:
\begin{equation}|2\rangle_{f}=\left(
                                \begin{array}{c}
                                 \frac{1}{\sqrt{2}}|u\bar{u}+d\bar{d}\rangle|c\bar{c}\rangle~~~~\text{for}~~~~~\omega J/\psi~~~\text{mesons} \\
                                 \frac{1}{\sqrt{2}}|u\bar{u}-d\bar{d}\rangle|c\bar{c}\rangle~~~~\text{for}~~~~~\rho J/\psi~~~~\text{mesons} \\
                                \end{array}
                              \right).
\label{e3.4}\end{equation}
This gives in eq.(\ref{e3.6})
\begin{equation}  _{f}\langle 1|2\rangle_{f}=\frac{1}{\sqrt{2}} \label{flavor overlap} \end{equation}
\noindent for both $\omega J/\psi$ and $\rho J/\psi$ in channel 2.

\section{Solving The Integral Equations}\label{solveintegraleqs}

When eqs. \eqref{e3.7a}-\eqref{e15} and eqs. \eqref{spin overlap},
\eqref{flavor overlap} are substituted in eq. (\ref{e3.6}), we get
the following equation
 \begin{eqnarray} \int
d^3\textbf{R}_{k}^{\prime}\Big[\textbf{K}_{k
k}(\textbf{R}_{k},\textbf{R}_{k}^{\prime})+ \textbf{V}_{k
k}(\textbf{R}_{k},\textbf{R}_{k}^{\prime})+(\sum_{i=1}^{\overline{4}}m_{i}-E_{c})
\textbf{N}_{kk}(\textbf{R}_{k},\textbf{R}_{k}^{\prime})\Big]\chi_{k}(\textbf{R}_{k}^{\prime})+\nonumber~~~~\\
\int_{l\neq
k}d^3\textbf{R}_{l}\Big[\textbf{K}_{kl}(\textbf{R}_{k},\textbf{R}_{l})+
\textbf{V}_{kl}(\textbf{R}_{k},\textbf{R}_{l})+(\sum_{i=1}^{\overline{4}}m_{i}-E_{c})
\textbf{N}_{kl}(\textbf{R}_{k},\textbf{R}_{l})\Big]\chi_{l}(\textbf{R}_{l})=0,~~~~~~\label{e3.13}
\end{eqnarray}

\noindent with the kernels
$\textbf{K}_{kl}(\textbf{R}_{k},\textbf{R}_{l}^{\prime})$,
$\textbf{V}_{kl}(\textbf{R}_{k},\textbf{R}_{l}^{\prime})$ and
$\textbf{N}_{kl}(\textbf{R}_{k},\textbf{R}_{l}^{\prime})$ defined,
in the notation of eq. (\ref{e3.6}), by
\begin{eqnarray} \int
d^3\textbf{y}_{k}d^3\textbf{z}_{k} \xi_{k}(\textbf{y}_{k})
\zeta_{{k}}(\textbf{z}_{k})K_{kl}\hspace{.02in}
\chi_{l}(\textbf{R}_{l})\xi_{l}(\textbf{y}_{l})
\zeta_{l}(\textbf{z}_{l})= \frac{2}{\delta_{kl}+1}\int
d^3\textbf{R}_{l}^{\prime}\textbf{K}_{kl}(\textbf{R}_{k},\textbf{R}_{l}^{\prime})
\chi_{l}(\textbf{R}_{l}^{\prime})~~~~~\label{e3.14}
\end{eqnarray}
\begin{eqnarray}
\int
d^3\textbf{y}_{k}d^3\textbf{z}_{k} \xi_{k}(\textbf{y}_{k})
\zeta_{{k}}(\textbf{z}_{k})V_{kl}\hspace{.02in}\chi_{l}(\textbf{R}_{l})\xi_{l}(\textbf{y}_{l})
\zeta_{l}(\textbf{z}_{l})=\frac{2}{\delta_{kl}+1}
\int
d^3\textbf{R}_{l}^{\prime}\textbf{V}_{kl}(\textbf{R}_{k},\textbf{R}_{l}^{\prime})
\chi_{l}(\textbf{R}_{l}^{\prime})~~~~\label{e3.15}
\end{eqnarray}
\begin{eqnarray}
\int d^3\textbf{y}_{k}d^3\textbf{z}_{k} \xi_{k}(\textbf{y}_{k})
\zeta_{{k}}(\textbf{z}_{k})N_{kl}\hspace{.02in}\chi_{l}(\textbf{R}_{l})\xi_{l}(\textbf{y}_{l})
\zeta_{l}(\textbf{z}_{l})=\frac{2}{\delta_{kl}+1} \int
d^3\textbf{R}_{l}^{\prime}\textbf{N}_{kl}(\textbf{R}_{k},\textbf{R}_{l}^{\prime})
\chi_{l}(\textbf{R}_{l}^{\prime}).~~~~\label{e3.16}\end{eqnarray}
\noindent The factor $\frac{2}{\delta_{kl}+1}$ takes care of the
off-diagonal spin and flavor overlap factors both
$=\frac{1}{\sqrt{2}}$.
The spatial integrations on the left hand side of eqs.
(\ref{e3.14}-\ref{e3.16}) and resulting kinetic energy, interaction
and normalization kernels are reported in Appendix \ref{app c}.
A comparison of kernels themselves can have a dynamical result; ref.
\cite{Masutani} tells that if the interaction kernel is proportional
to the normalization kernel, the interaction does not contribute to
the interaction between mesons. Eqs.~(\ref{Nkk}) and (\ref{e3.24})
in the Appendix A show that such is the case in our calculations for
a single channel completely described by the diagonal terms in
kernels in these equations. For quadratic confinement in one channel
approximation ref. \cite{Masutani} also gets the same result for the
interaction between the mesons. But with an improved model for two
channel meson-meson interaction our full results are obtained by
substituting diagonal as well as off-diagonal terms in eq.
(\ref{e3.13}) and in our case the interaction kernel is not
proportional to the normal kernel and hence the quadratic
confinement contributes to the interaction between mesons. This is a
non-trivial result that can be compared with the baryon-baryon
interaction where refs. \cite{OkaYazaki 1981, Burger} report the
quark-exchange kernel generated by purely quadratic confinement
being proportional to the norm kernel and thus in this case the
quadratic confinement does not contribute to (the baryon baryon)
interaction. If confinement contributes to the meson-meson
interaction, it may worsen the van der Walls force problem between
isolated mesons that results by a sum of two-body potential but is
against the empirical evidence. But, as mentioned in the
introduction, we are finding meson level dynamical implications of
the quark potential model improvements~\cite{P. Pennanen, green2, A.
M. Green and P. Pennanen} that use multi-quark interactions in form
of the $f$ factor to avoid this problem; many works, including ref.
\cite{OkaYazaki} closely related to \cite{OkaYazaki 1981}, had
earlier suggested that many body interaction is needed to avoid this
long range interaction between mesons.

Using all the kernels, we get two integral equations for $k=1,2$; we
write here one of them:
\begin{eqnarray}
\Big[\frac{3}{4}(\omega_{21}+\omega_{22})-\frac{s_{2}}{2m}\underline{\nabla}_{\textbf{R}_{2}}^{2}-\frac{8}{3}\bar{C}-
4C[d_{21}^{2}+d_{22}^{2}]+2m(r+1)-E_{c}\Big]\chi_{2}(\textbf{R}_{2})\nonumber\\
+l_{0} \int
d^3\textbf{R}_{1}\Bigg[-\frac{1}{2m}\frac{1}{6}\Big[r_{21}\textbf{R}_{1}^{2}+r_{22}\textbf{R}_{2}^{2}+r_{20}\Big]+
\frac{1}{2}\Big[n_{1}\textbf{R}_{1}^{2}+n_{0}\Big]\nonumber\\-\frac{1}{6}\Big(E_{c}+\frac{8}{3}\bar{C}-2m(r+1)\Big)
\Bigg]\text{exp}(-l_{1}\textbf{R}_{1}^{2}-l_{2}\textbf{R}_{2}^{2})\chi_{1}(\textbf{R}_{1})=0.\label{e3.80}\end{eqnarray}
Here $s_{2}$, $\omega^,s$, $l^,s$, $n^,s$ and $r^,s$ depend upon the
constituent quark masses, sizes of mesons the parameter $k_f$ and
$b_s$; see Appendix \ref{app c}.
It is clear from this equation that off-diagonal parts vanish for
large values of $\textbf{R}_{1}$ and $\textbf{R}_{2}$. With no
interaction in this limit between the two mesons, the total center
of mass energy in the large separation limit will be the sum of
kinetic energies of the relative motion of mesons and masses of the
two mesons. This gives an alternative mesonic form for the diagonal
terms survived in the large distance (no interaction limit), which can
be utilized to write our integral equations as
\begin{eqnarray}
\Big[M_{x}+M_{J/\psi}-\frac{1}{2 \mu_{x
J/\psi}}\underline{\nabla}_{\textbf{R}_{2}}^{2}-E_{c}\Big]\chi_{2}(\textbf{R}_{2})+
l_{0} \int
d^3\textbf{R}_{1}\Bigg[-\frac{1}{2m}\frac{1}{6}\Big[r_{21}\textbf{R}_{1}^{2}+r_{22}\textbf{R}_{2}^{2}+r_{20}\Big]\nonumber\\+
\frac{1}{2}\Big[n_{1}\textbf{R}_{1}^{2}+n_{0}\Big]-\frac{1}{6}\Big(E_{c}+\frac{8}{3}\bar{C}-2m(r+1)\Big)
\Bigg]\text{exp}(-l_{1}\textbf{R}_{1}^{2}-l_{2}\textbf{R}_{2}^{2})\chi_{1}(\textbf{R}_{1})=0,\hspace{1cm}\label{e3.86}\end{eqnarray}
with $x=\omega, \rho$, and a similar one with the diagonal term as $\Bigg[M_{D}+M_{\bar{D}^{0*}}-\frac{1}{2
\mu_{\bar{D}^{0}\bar{D}^{0*}}}\underline{\nabla}_{\textbf{R}_{1}}^{2}-E_{c}\Bigg]$.
By taking Fourier transform of eq. (\ref{e3.86}), we get
\begin{eqnarray}
&&\Big[M_{x}+M_{J/\psi}+\frac{1}{2 \mu_{x
J/\psi}}\textbf{P}_{2}^{2}-E_{c}\Big]\chi_{2}(\textbf{P}_{2})
-\frac{1}{2m}\frac{r_{22}}{6}A_{1}(l_{1})F_{b}(\textbf{P}_{2},l_{2})\nonumber\\&&+\Big[\Big(-\frac{1}{2m}\frac{r_{20}}{6}+
\frac{n_{0}}{2}-\frac{E_{c}^{\prime}}{6}\Big)A_{1}(l_{1})+\Big(-\frac{1}{2m}\frac{r_{21}}{6}+\frac{n_{1}}{2}\Big)B_{1}(l_{1})\Big]
F_{a}(\textbf{P}_{2},l_{2})=0.\hspace{1cm}\label{e3.88}\end{eqnarray}
where, $ E_{c}^{\prime}=E_{c}+\frac{8}{3}\bar{C}-2m(r+1). $
In these equations
\begin{eqnarray}
A_{k}(u)=l_{0}\int
d^{3}\textbf{R}_{k}\text{exp}[-u\textbf{R}_{k}^{2}]\chi_{k}(\textbf{R}_{k})\label{e3.91}\end{eqnarray}
\begin{eqnarray}
B_{k}(u)=l_{0}\int
d^{3}\textbf{R}_{k}\text{exp}[-u\textbf{R}_{k}^{2}]\textbf{R}_{k}^{2}\chi_{k}(\textbf{R}_{k})\label{e3.92}\end{eqnarray}
\begin{eqnarray}
F_{a}(\textbf{P}_{k},u)\equiv\int
\frac{d^{3}\textbf{R}_{k}}{(2\pi)^{\frac{3}{2}}}\text{exp}[i\textbf{P}_{k}.\textbf{R}_{k}]\text{exp}[-u\textbf{R}_{k}^{2}]=\frac{1}{(2u)^{\frac{3}{2}}}
\text{exp}\Bigg[-\frac{\textbf{P}_{k}^{2}}{4u}\Bigg]\label{e3.93}\end{eqnarray}
\begin{eqnarray}
F_{b}(\textbf{P}_{k},u)\equiv\int
\frac{d^{3}\textbf{R}_{k}}{(2\pi)^{\frac{3}{2}}}\text{exp}[i\textbf{P}_{k}.\textbf{R}_{k}]\textbf{R}_{k}^{2}\text{exp}[-u\textbf{R}_{k}^{2}]=F_{a}(\textbf{P}_{k},u)\Bigg[\frac{1}{2u}\Bigg]\Bigg[3-\frac{\textbf{P}_{k}^{2}}{2u}\Bigg].\hspace{.5in}\label{e3.94}
\end{eqnarray}
For the incoming waves in the first channel, our two integral equations (eq. (\ref{e3.88}) and the other one; we \emph{now} write both)
can be formally solved \cite{B. Masud} as (see appendix-\ref{app d} for details)
\begin{eqnarray}
\chi_{1}(p_{1})=\frac{\delta(p_{1}-p_{c}(1))}{p_{c}^{2}(1)}-
\frac{1}{\Delta_{1}(p_{1})}\Big[W_{1}^{(1)}A_{2}(l_{2})+W_{2}^{(1)}B_{2}(l_{2})\Big]\hspace{.5in}\label{e3.125}
\end{eqnarray}
\begin{eqnarray}
\chi_{2}(p_{2})=-
\frac{1}{\Delta_{2}(p_{2})}\Big[W_{1}^{(2)}A_{1}(l_{1})+W_{2}^{(2)}B_{1}(l_{1})\Big].\label{e3.126}
\end{eqnarray}
Here
\begin{eqnarray}
\Delta_{1}(p_{1})=\frac{p_{1}^{2}}{2
\mu_{D^{0}\bar{D}^{0*}}}+M_{D^0}+M_{\bar{D}^{0*}}-E_{c}-i \epsilon
\texttt{ }\label{e3.99}
\end{eqnarray}
for an infinitesimal $\epsilon$. Similarly,
\begin{eqnarray}
\Delta_{2}(p_{2})=\frac{p_{2}^{2}}{2 \mu_{x
J/\psi}}+M_{x}+M_{J/\psi}-E_{c}-i \epsilon \texttt{
}\label{e3.100}\end{eqnarray}
\begin{eqnarray}
p_{c}(1)=
\sqrt{2\mu_{D^{0}\bar{D}^{0*}}(E_{c}-M_{D^0}-M_{\bar{D}^{0*}})}\hspace{.5in}\label{e3.101}
\end{eqnarray}
\begin{eqnarray}
p_{c}(2)= \sqrt{2 \mu_{x
J/\psi}(E_{c}-M_{x}-M_{J/\psi})}.\label{e3.102}
\end{eqnarray}
\begin{eqnarray}
W_{1}^{(1)}=\Big[-\frac{1}{2m}\frac{r_{11}}{6}+\frac{n_{1}}{2}\Big]F_{b}(p_{c}(1),l_{1})+
\Big[-\frac{1}{2m}\frac{r_{10}}{6}+\frac{n_{0}}{2}-\frac{E_{c}^{\prime}}{6}\Big]F_{a}(p_{c}(1),l_{1})\label{e3.127}
\end{eqnarray}
\begin{eqnarray}
W_{2}^{(1)}=-\frac{1}{2m}\frac{r_{12}}{6}F_{a}(p_{c}(1),l_{1})\label{e3.128}
\end{eqnarray}
\begin{eqnarray}
W_{1}^{(2)}=-\frac{1}{2m}\frac{r_{22}}{6}F_{b}(p_{c}(2),l_{2})+
\Big[-\frac{1}{2m}\frac{r_{20}}{6}+\frac{n_{0}}{2}-\frac{E_{c}^{\prime}}{6}\Big]F_{a}(p_{c}(2),l_{2})\hspace{.5in}\label{e3.129}
\end{eqnarray}
\begin{eqnarray}
W_{2}^{(2)}=\Big[-\frac{1}{2m}\frac{r_{21}}{6}+\frac{n_{1}}{2}\Big]F_{a}(p_{c}(2),l_{2}).\label{e3.130}
\end{eqnarray}
From eqs. (\ref{e3.125}) and (\ref{e3.126}) we can read off the
T-matrix elements $T_{11}$ and $T_{21}$ \cite{B. Masud} as co-efficient of Green's function operators
$-\frac{1}{\Delta_{1}(p_{1})}$ and $-\frac{1}{\Delta_{2}(p_{2})}$
respectively. So, we have
\begin{eqnarray}
T_{11}=2\mu_{D^{0}\bar{D}^{0*}}\frac{\pi}{2}p_{c}(1)\Big[W_{1}^{(1)}A_{2}(l_{2})+W_{2}^{(1)}B_{2}(l_{2})\Big]\label{e3.132}
\end{eqnarray}
\begin{eqnarray}
T_{21}=2\mu_{x
J\psi}\frac{\pi}{2}p_{c}(1)\sqrt{\frac{v_{2}}{v_{1}}}\Big[W_{1}^{(2)}A_{1}(l_{1})+W_{2}^{(2)}B_{1}(l_{1})\Big],\label{e3.133}
\end{eqnarray}
where $v_{1}=p_{c}(1)/\mu_{D^{0}\bar{D}^{0*}}~\text{and}~v_{2}=p_{c}(2)/\mu_{xJ\psi}\texttt{ }.$
 Similarly $T_{22}$ and $T_{12}$ can be found for the incoming waves in the 2nd channel, with the $V_2$  in Appendix-\ref{app d} accordingly changed. These are
\begin{eqnarray}
T_{22}=2\mu_{x
J\psi}\frac{\pi}{2}p_{c}(2)\Big[W_{1}^{(2)}A_{1}(l_{1})+W_{2}^{(2)}B_{1}(l_{1})\Big]\label{e3.135}
\end{eqnarray}
\begin{eqnarray}
T_{12}=2\mu_{D^{0}\bar{D}^{0*}}\frac{\pi}{2}p_{c}(2)\sqrt{\frac{v_{1}}{v_{2}}}\Big[W_{1}^{(1)}A_{2}(l_{2})+W_{2}^{(1)}B_{2}(l_{2})\Big].\label{e3.136}
\end{eqnarray}
\section{Parameters fixing}\label{fitparameters}
At the quark level we adopt the model of refs. \cite{P. Pennanen,green2} that includes the parameters $k_f$ and $b_s$
in the gluonic field overlap factor $f$. We take the
 value of $k_{f}=0.075$ \cite{green2} and $b_{s}$ as $0.18$ Ge$V^{2}$ \cite{Fumiko Okiharu}.
  Our own contribution is in using the meson wave functions to find the hadron level implications for our chosen channels. These are eigenfunctions of
   potential of eq. (\ref{e16}) which has parameters $C$ and $\bar{C}$ whose numerical values we find by equating relevant terms in the large distance limit
   of eq. (\ref{e3.80}) to the $J/\psi$ meson mass; see eq. (\ref{e3.86}). This gives
\begin{eqnarray}M_{J/\psi}=\frac{3}{4}\omega_{22}- \frac{4}{3}\bar{C}-
4Cd_{22}^{2}+2m_{c}.\label{e3.137}
\end{eqnarray}
Comparing eqs. (\ref{e3.7b}) and (\ref{e16}) with the
standard form of potential of a simple harmonic oscillator gives
$-4C/3=\mu_{c\bar{c}} \omega_{22}^{2}/2$. Using this and $\omega_{22}=1/m_{c}d_{22}^{2}$, we can eliminate $C$ and the size $d_{22}$ in favor of $\omega_{22}$
to get
\begin{eqnarray}M_{J/\psi}=\frac{3}{2}\omega_{22}- \frac{4}{3}\bar{C}+2m_{c}.\label{e3.143}
\end{eqnarray}

It is to be noted that this equation tells that in our model the dynamics of quarks, incorporating the effects of the glounic field in the form of potential,
causes the mass of the quark antiquark cluster (a meson) to be
a few percent different to the mere sum $2m_c$ of quark masses. Our choice in eq. (\ref{e16}) of using a simple harmonic oscillator potential with a known total
energy allows us to write kinetic energy as known total energy minus potential energy. Thus the origin of clustering,
or charm-anticharm quarks binding, is in the parameters $C$ and $\bar C$ of the potential in eq. (\ref{e16}). The factor $- \frac{4}{3}$ in eq. (\ref{e3.143})
multiplying $\bar C$ is a color factor which is the color expectation value of the $\mathbf{F}_{i}.\mathbf{F}_{j}$ operator in eq. (\ref{e9}) and we have
defined $C$ by $-4C/3=\mu_{c\bar{c}} \omega_{22}^{2}/2$ with positive $\omega_{22}$, making $C$ to be negative. Below we replace $C$ by $\omega_{22}$ as
our model parameter.

It is to be noted that there is no spin dependence in this modeled origin of the quark-antiquark clustering or binding; our neglect of hyperfine interaction
is responsible for this spin-independence. Thus, we do not make separate models of two different spin states of otherwise one quark-antiquark clustering of,
say, a specified angular momentum between a quark and an antiquark. Specifically, this means that we are not able to model the mass difference of $J/\psi(1S)$
and $\eta_{c}$ which have the same quark antiquark angular momentum $L=0$ and differ only in spin dependence. Thus we fit our remaining parameters $\omega_{22}$
and $\bar{C}$, mentioned in the above paragraph, to the spin averaged masses of charmonium in the state $1S$ and the state $2S$. This
replaces eq.(\ref{e3.143}) by
\begin{eqnarray}\frac{3M_{J/\psi}(1S)+M_{\eta_{c}}(1S)}{4}=\frac{3}{2}\omega_{22}- \frac{4}{3}\bar{C}+2m_{c}.\label{e3.138}
\end{eqnarray}
For a comparison, ref.~\cite{spinaverage} uses spin averaged ${\bar
b}b$ spectrum in its Fig. 1. An explicit formula for spin averaged
mass can be seen as eq. (3.1) of  ref.~\cite{thesisspinaverage}.

And for 2S state 3/2 is replaced by 7/2 because of 3-d S.H.O.
$E_{nlm}=\omega_{22}(4n+2l+3)/2$
\cite{Nouredine Zettili}, for this
$n=1$ and $l=0$. The corresponding equation is
\begin{eqnarray}\frac{3M_{\psi}(2S)+M_{\eta_{c}}(2S)}{4}=\frac{7}{2}\omega_{22}- \frac{4}{3}\bar{C}+2m_{c}.\label{e3.139}
\end{eqnarray}
\noindent Put the values of masses $M_{J/\psi}(1S)=3.0969$ GeV,
$M_{\eta_{c}}(1S)=2.9803$ GeV, $M_{\psi}(2S)=3.6861$ GeV and
$M_{\eta_{c}}(2S)=3.6370$ GeV from (PDG) ref. \cite{K Nakamura} in
eqs. (\ref{e3.138}) and (\ref{e3.139}) and solving them
simultaneously, we get $\bar{C}=0.2592$ GeV and $\omega_{22}=0.3030$ GeV
for a charm-anticharm cluster; we use the constituent quark masses values
$m_{c}=1.4794$ GeV and $m=0.33$ GeV (for light quarks) of ref.~\cite{ES Swanson}.
For angular frequencies $\omega's$ and hence sizes of heavy-light and light-light clusters, we used the S.H.O. property that size square is inversely
proportional to the square root of the relevant reduced mass (that is of quark and antiquark in the meson).

\section{Results and Conclusion}\label{sec5}

According to eqs. \eqref{e3.132}, \eqref{e3.133}, \eqref{e3.135} and \eqref{e3.136}, the $T$-matrix elements
are given in terms of the elements of $V_1$ and $V_2$ column matrices which satisfy the inhomogeneous eq. \eqref{e3.120}.  These solutions of the eq. (\ref{e3.120})
are finite if $\texttt{det}W\neq0$. Using the numerical values of our parameters, we calculate the $T$ matrix elements as a function of energy which in turn give
the spin averaged cross-sections using the following relation \cite{Steven Weinberg}
\begin{eqnarray}
\sigma_{ii^\prime}=\frac{4
\pi}{p_{c}^{2}(i^\prime)}\sum_{J}\frac{(2J+1)}{(2s_{1}+1)(2s_{2}+1)}|
T_{ii^\prime}|^{2},\label{e3.140}
\end{eqnarray}
\noindent where $J$ is the total angular momentum of the mesons and $s_{1}$ and $s_{2}$ are the spin of the two
incoming mesons. (For the definition of $p_{c}^{2}(i^\prime)$, see eqs.
\eqref{e3.101} and \eqref{e3.102} above.) Here $i,i^\prime=1,2$
label our channels. In
fig. 3 we show spin averaged cross sections
versus  $T_{c}=E_{c}-M_{\bar{D}^{0}}-M_{D^{0*}}$ for the process $\bar{D}^{0}D^{0*}\longrightarrow \bar{D}^{0}D^{0*}$ and $T_{c}=E_{c}-M_{\omega}-M_{J/\psi}$
for the processes $\bar{D}^{0}D^{0*}\longrightarrow \omega J/\psi$, $\omega
J/\psi \longrightarrow \bar{D}^{0}D^{0*}$ and $\omega J/\psi
\longrightarrow \omega J/\psi$ for the QCD-based model that we are using, which means the parameter $k_f$ is taken 0.075.
The cross sections are smooth (without any  peak),
relatively small and decrease very rapidly with $T_{c}$.
In fig. 4 the cross sections of the same processes are given for the sum of two-body potential model, that is setting the value of the parameter $k_{f}$ as zero.
The cross sections in this case are smooth, relatively large and again decrease rapidly with $T_{c}$.
To find the cross sections of the processes given in fig. 3 or 4, we assume that the channel 1 and 2 are $\bar{D}^{0}D^{0*}$ and $\omega J/\psi$ respectively.
However, if the channel 2 is taken $\rho J/\psi$ then we can obtain the cross sections of the processes $\bar{D}^{0}D^{0*}\longrightarrow \bar{D}^{0}D^{0*}$,
$\bar{D}^{0}D^{0*}\longrightarrow \rho J/\psi$, $\rho J/\psi \longrightarrow \bar{D}^{0}D^{0*}$ and $\rho J/\psi
\longrightarrow \rho J/\psi$, here $T_{c}=E_{c}-M_{\rho}-M_{J/\psi}$ for all the processes excluding the process
$\bar{D}^{0}D^{0*}\longrightarrow \bar{D}^{0}D^{0*}$ where we have taken $T_{c}=E_{c}-M_{\bar{D}^{0}}-M_{D^{0*}}$.
The plots of these cross sections are given in fig. 5 and 6 for $k_f=0.075$ and 0 respectively. We again find that the cross sections are
suppressed when Gaussian $f$ factor is included. It is noted that the first process $\bar{D}^{0}D^{0*}\longrightarrow \bar{D}^{0}D^{0*}$,
which is common in both sets of processes,
was checked to have the same cross section whereas the values of cross sections of other processes are somewhat different.

\begin{figure}[!h]
\includegraphics[scale=1,angle=-0]{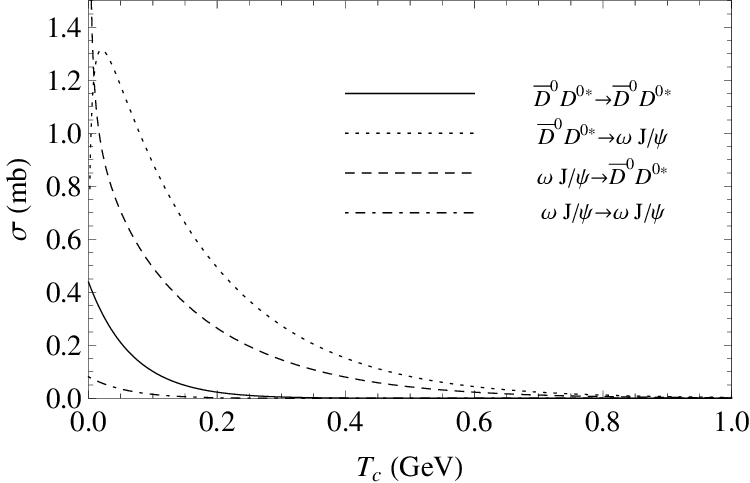} \caption{Total spin averaged cross
sections for Gaussian form of \emph{f} with $k_{f}=0.075$ versus $T_{c}$ when channel 2 is taken $\omega J/\psi$.} \label{graph6}
\end{figure}

\begin{figure}[!h]
\includegraphics[scale=1,angle=-0]{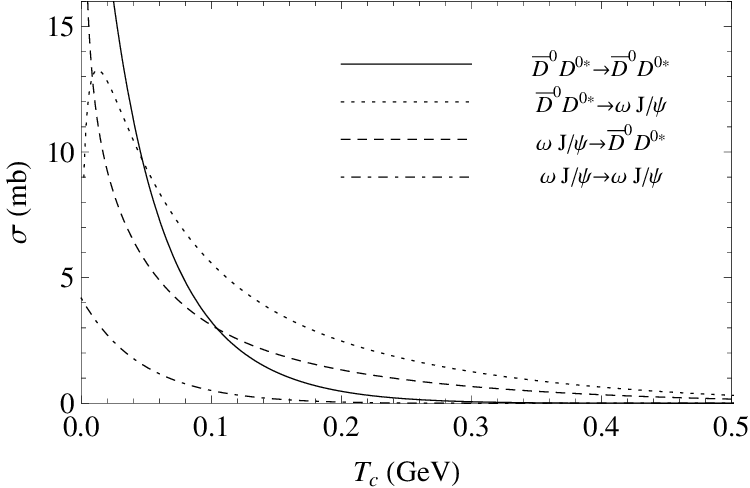} \caption{Total spin
averaged cross sections for $k_{f}=0$ versus $T_{c}$ when channel 2 is taken $\omega J/\psi$.} \label{graph14}
\end{figure}

\begin{figure}[!h]
\includegraphics[scale=1,angle=-0]{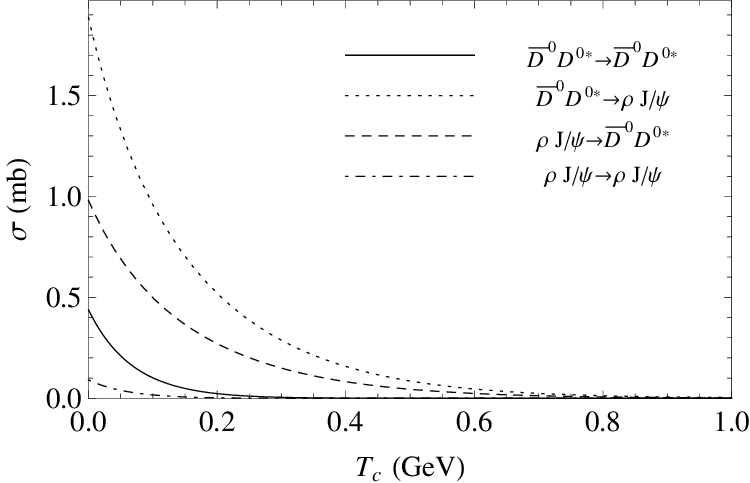} \caption{Total spin
averaged cross sections for Gaussian form of \emph{f} with $k_{f}=0.075$ versus $T_{c}$ when channel 2 is taken $\rho J/\psi$.}
\label{graph15}
\end{figure}

\begin{figure}[!h]
\includegraphics[scale=1,angle=-0]{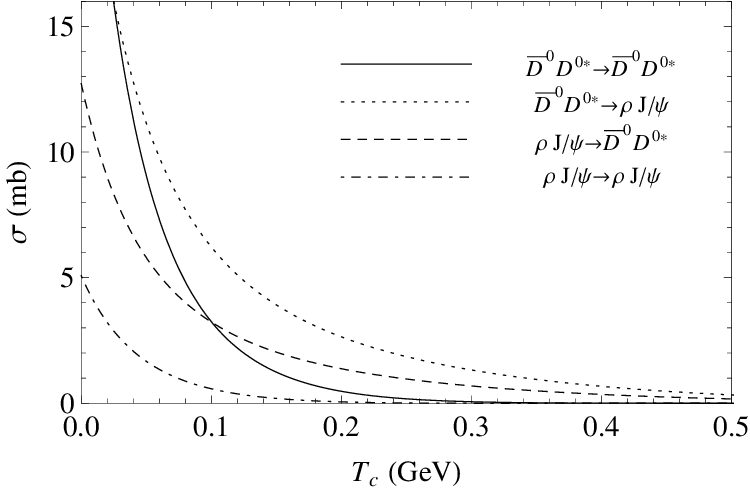} \caption{Total spin
averaged cross sections for $k_{f}=0$ versus $T_{c}$ when channel 2 is taken $\rho J/\psi$.} \label{graph15p}
\end{figure}

At $\det{W}=0$ the solution of eq. (\ref{e3.120}) diverges, which corresponds to a pole of scattering amplitude and represents a bound state (resonance)
with respect to a given process if its energy is less (greater) than the process threshold which is equal to total rest mass of the final (inital)
particles in case of endothermic (exothermic) processes respectively. In order to calculate the energy where the pole exist for our $q^{2}\bar{q}^{2}$ system
we simply have to solve $\det{W}=0$ for the energy variable.
We find that $\texttt{det}W\neq0$ for all $T_{c}>0$
 when $k_{f}=0$ and $k_{f}=0.075$.
 These results are consistent with the
plots in figs. \ref{graph6}-\ref{graph15p}
of the cross sections in which no resonating peak appears for these values of  $k_{f}$.

 As refs. \cite{Swanson 04,Braaten a5, Braaten a7, P. Colangelo and F. De Fazio and S. Nicotri, Masayasu Harada and Yong-Liang Ma, T. Mehen and R. Springer}
 have pointed out that $\bar{D}^{0}D^{0*}$ may form a bound state, it is worth examining if by changing the strength of our interaction
 we can get a meson-meson bound state or resonance.
 To do this analysis we introduce a parameter $I_{0}$
as in ref. \cite{masud} changing the net strength of our meson-meson
interaction. Physically, this parameter $I_{0}$ tells how far we are from getting a bound state at
 $3872$ MeV if we study only one component
  $\bar{D}^{0}D^{0*}$ of the full exotic meson $X(3872)$ along with using other approximations. Any deviation of $I_{0}$ from 1 suggests how much can
  we improve modeling of this exotic meson. We implemented this re-scaling of the interaction
strength by multiplying the off-diagonal terms of our potential,
kinetic energy, and normalization matrices (i.e., multiplying
$l_{0}$ of eq. \ref{e3.80} and the other coupled integral equation by $I_{0}$). A value of
$I_{0}$ away from 1 (for all the above results) changes the energy
where condition $\det{W}=0$ is satisfied. Energy of the bound state
generally depends upon strength parameter
$I_{0}$ of the interaction in two
possible ways \cite{Taylor}; either the energy of the bound state
increases or decreases with the strength parameter. In the former
case it is usually called virtual state whereas in later case we
give it the name of proper bound state. In fig. 7 we show the
dependence of the c.m. energy  at pole on the strength parameter
$I_{0}$ subject to the constraint $\text{det} W=0$ by different
curves for $k_{f}=0$, $0.05$, $0.075$, and $0.1$ respectively. While
solving $\det{W}=0$ we note that the solution can be obtained
conveniently if we put the value of $E_{c}$ and other kinematical
variables and solve it for $I_{0}$ rather than solving it for
$E_{c}$. In this way we find that the resultant equation is
quadratic in $I_{0}$, which means we may have two values of $I_{0}$
corresponding to one value of $E_{c}$. However, we find that one of
two roots is always complex and real root is found to be continuous
function of $E_{c}$ as is indicated by the continuous curves in the
fig. 7, in which solid and dashed segments corresponds to first and
second real root respectively. These curves show that corresponding
to each $k_{f}$, the resonance energy $E_{c}$ increases with $I_{0}$
provided that $I_{0}$ is greater than a critical value, which depend
on the value of $k_{f}$. For example for $k_{f}=0.075$ the critical
$I_{0}= 2.89$ for 2nd-channel being $\omega J/\psi$. It means that
pole of the scattering amplitude does not exist at $I_{0}<2.89$ when
$f$ factor is included at $k_{f}=0.075$. Similarly for $k_{f}=0$ the
critical $I_{0}=1.38$. This explains why there appears no resonating
peak in the plots of the cross sections when $I_{0}$ is taken 1
irrespective of the value of $k_{f}$. The curves given in fig. 7 are
produced by assuming that the channel 1 and 2 are
$\bar{D}^{0}D^{0*}$ and $\omega J/\psi$ respectively. We find
similar results when the channel 2 is taken $\rho J/\psi$, as shown
in fig. 8. In table \ref{s1} we give the critical values of $I_{0}$
corresponding to different values of $k_{f}$ for the two choices of
channel 2. It is also noted that minimum $E_{c}$ at which
$\det{W}=0$ is 3.881 and 3.872 GeV for channel 2 being $\omega
J/\psi$ and $\rho J/\psi$ respectively irrespective of the value of
$k_{f}$. These values are slightly greater or equal to
$m_{D^{0}}+m_{D^{*0}}$ (3.872 GeV), $m_{\omega}+m_{J/\psi}$ (3.88
GeV), and $m_{\rho}+m_{J/\psi}$ (3.872 GeV). This implies that in
our case pole  of scattering amplitude corresponds to a resonance in
the system. Thus, we conclude that $c\bar{c}u\bar{u}$ system cannot
resonate whether we assume sum of two-body approach (i.e.,
$k_{f}=0$) or include QCD effect in terms of gluonic field overlap
factor $f$ at $I_{0}=1$. However, the resonance may be produced if
the interaction strength
$I_{0}$ is increased at least by the factor of 1.38
(1.35) and 2.89 (2.82) for $k_{f}=0$ and 0.075 respectively when
channel 2 is $\omega J/\psi$ ($\rho J/\psi$).
It is tempting to associate the resonance in $q^2\bar{q}^2$ with $\bar{D}^{0}D^{0*}$ component of X(3872). The result that this resonance appears only
when interaction strength parameter $I_{0}$ is greater than a critical value may be related with the use of various approximations used in this work
including ignoring the annihilation effects of light quark flavors and using quadratic confinement. As for the full X(3872), our neglect of its $c\bar{c}$
component  \cite{y Dong and a Faessler,Ortega and Segovia PRD, Bugg, Eitchen Lane, Voloshin 05,Lee I W and Faessler A and
Gutsche T and Lyubovitskij V E,Kalashnikova Yu S and Nefediev A V}
may also be responsible for deviation of the parameter $I_{0}$ away from 1.
If future improvements beyond our approximations are equivalent to an effective $I_{0}$ that is lesser than one, our work would imply that $\bar{D}^{0}D^{0*}$
do not form a bound state and hence there can not be a role of $\bar{D}^{0}D^{0*}$ molecule in the structure of X(3872).
If the resulting effective $I_{0}$ is increased beyond the critical values mentioned in table \ref{s1},
the $\bar{D}^{0}D^{0*}$ bound state may represent X(3872).

\begin{figure}[!h]
\includegraphics[scale=1,angle=-0]{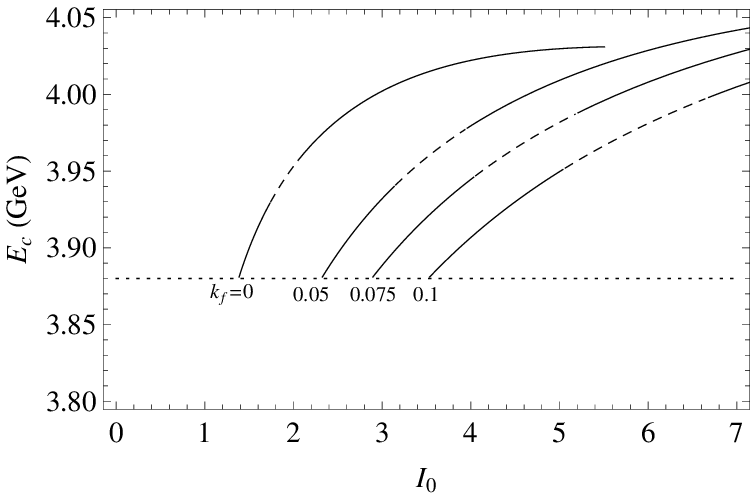} \caption{Total centre of mass energy at pole verses strength parameter $I_{0}$, for different values of $k_{f}$, for 2nd channel being $\omega J/\Psi$.}
\label{graph255}
\end{figure}

\begin{figure}[!h]
\includegraphics[scale=1,angle=-0]{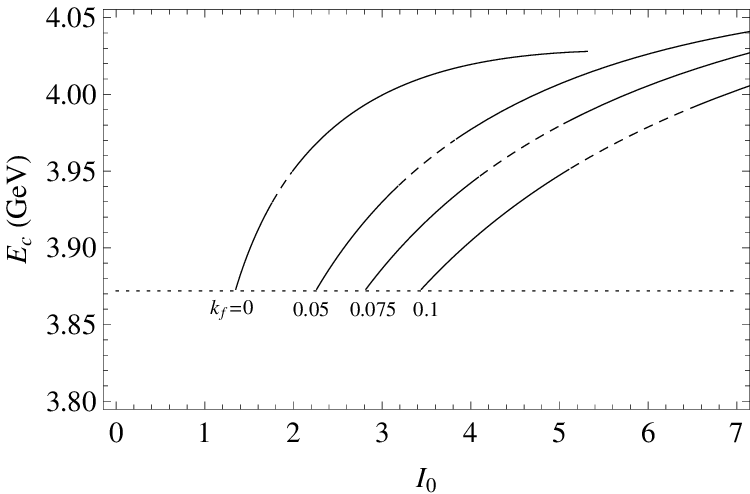} \caption{Total centre of mass energy at pole verses strength parameter $I_{0}$, for different values of $k_{f}$, for 2nd channel being $\rho J/\Psi$.}
\label{graph255}
\end{figure}

\begin{table}[tbp]
\begin{center}
\renewcommand{\arraystretch}{1.3}
\begin{tabular}{|c|c|c|}
  \hline
     & Channel-2 ($\omega J/\psi$) & Channel-2 ($\rho J/\psi$) \\ \hline
  $k_{f}$ & Critical $I_{0}$ & Critical $I_{0}$ \\ \hline
  0 & 1.3863 & 1.3487 \\
   0.05 & 2.3253 & 2.2610 \\
   0.075 & 2.8950 & 2.8164 \\
   0.1 & 3.5357 & 3.4422 \\
  \hline
\end{tabular}
\caption{Critical $I_{0}$ for different values of $k_{f}$.}\label{s1}
\end{center}
\end{table}~~~~~~~~~~~~~~
\section{Conflicts of Interest}
The authors declare that there is no conflict of interest regarding the publication of this manuscript.

\begin{appendix}
\section{}\label{app c}
Here is told how we performed the spatial integrations on the left
hand side of eqs. (\ref{e3.14}-\ref{e3.16}) to read our kernels.
From figs. (\ref{fig1}) and (\ref{fig2}) we see that
$\textbf{y}_{1}, \textbf{z}_{1}$, $\textbf{R}_{1}$ and
$\textbf{y}_{2}, \textbf{z}_{2}$,
 $\textbf{R}_{2}$ form two linearly independent sets.
Thus for the diagonal terms $k=l$ in eq. (\ref{e3.6}),
$\chi_{l}(\textbf{R}_{l})$ can be taken out side of integration on
RHS of eq. (\ref{e3.16}). Thus normalization of
$\xi_{k}(\textbf{y}_{k})$, defined in eq.~(\ref{Gaussian1}) and a
similar $\zeta_{k}(\textbf{z}_{k})$, gives
\begin{equation}
\int
d^3\textbf{R}_{k}^{\prime}\textbf{N}_{kk}(\textbf{R}_{k},\textbf{R}_{k}^{\prime})
\chi_{k}(\textbf{R}_{k}^{\prime})=\chi_{k}(\textbf{R}_{k})\hspace{.5in}
\text{or} \end{equation}
\begin{equation}
\textbf{N}_{kk}(\textbf{R}_{k},\textbf{R}_{k}^{\prime})=\delta(\textbf{R}_{k},\textbf{R}_{k}^{\prime})
\label{Nkk}.
\end{equation}
For kinetic energy, in eq. (\ref{e15}) we can write for $k=1$ or
$k=2$
\begin{equation}\Big(\sum_{i=1}^{\overline{4}}-\frac{\underline{\nabla}_{i}^{2}}{2m_{i}}\Big)=-\frac{1}{2m}
[s_{k}\underline{\nabla}_{\textbf{R}_{k}}^{2}+q_{k}\underline{\nabla}_{\textbf{y}_{k}}^{2}+t_{k}\underline{\nabla}_{\textbf{z}_{k}}^{2}],
\label{e3.20}\end{equation} with $m$ the constituent mass of the
light quark, up or down and
\begin{equation}s_{1}=\frac{2}{r+1},\texttt{
} q_{1}=t_{1}=\frac{r+1}{r},\texttt{ } s_{2}=\frac{r+1}{2r},\texttt{
} q_{2}=2,\texttt{ } t_{2}=\frac{2}{r}.\end{equation} By using eq.
(\ref{e3.20}) in eq. \eqref{e3.14} and doing the required space
differentiations and integrations, we get
\begin{eqnarray}\textbf{K}_{kk}(\textbf{R}_{k},\textbf{R}_{k}^{\prime})=\delta(\textbf{R}_{k},\textbf{R}_{k}^{\prime})
\Big[\frac{3}{4}(\omega_{k1}+\omega_{k2})-\frac{s_{k}}{2m}\underline{\nabla}_{\textbf{R}_{k}}^{2}\Big]
\hspace{.5in} \text{with}\label{e3.21}\end{eqnarray}
\begin{eqnarray}\omega_{k1}=\frac{q_{k}}{2md_{k1}^{2}}
\text{  and  }
\omega_{k2}=\frac{t_{k}}{2md_{k2}^{2}}.\label{e3.22}\end{eqnarray}
For the potential energy matrix, by using eqs. \eqref{e16} and
\eqref{e3.7b} we get
\begin{equation}
V_{kk}=-\frac{4}{3}\Big[2\bar{C}+C\textbf{y}_{k}^{2}+C\textbf{z}_{k}^{2}\Big]
\label{Vkk}.
\end{equation}
 Using this in eq. \eqref{e3.15}
and doing the required integrations, we get
\begin{eqnarray}
\textbf{V}_{kk}(\textbf{R}_{k},\textbf{R}_{k}^{\prime})=\delta(\textbf{R}_{k},\textbf{R}_{k}^{\prime})\Big[-\frac{8}{3}\bar{C}-4C[d_{k1}^{2}+d_{k2}^{2}]\Big].\label{e3.24}
\end{eqnarray}
 Now for the off-diagonal elements
we have to replace $\textbf{y}_{1}$ and $\textbf{z}_{1}$ by
$\textbf{R}_{2}$ and $\textbf{g}_{1}$, where
\begin{eqnarray}
\textbf{g}_{1}=\textbf{y}_{1}+\textbf{z}_{1}.\label{e3.26}
\end{eqnarray}
 Only $\textbf{g}_{1}$ is
integrated. The rest is a function of $\textbf{R}_{2}$ and
$\textbf{R}_{1}$ (constant in this integration). Similarly we
replace $\textbf{y}_{2}$ and $\textbf{z}_{2}$ by $\textbf{R}_{1}$
and $\textbf{g}_{2}$, where
\begin{eqnarray}
\textbf{g}_{2}=\textbf{y}_{2}+r\textbf{z}_{2}.\label{e3.28}
\end{eqnarray}
 Only $\textbf{g}_{2}$ is
integrated. The rest is a function of $\textbf{R}_{1}$ and
$\textbf{R}_{2}$ (constant in this integration). We get from eqs.
\eqref{e3.7a}, \eqref{e7}, \eqref{e3.16} after doing all the
integrations other than $\textbf{R}_{l}$
\begin{eqnarray}
\textbf{N}_{12}(\textbf{R}_{1},\textbf{R}_{2})=\textbf{N}_{21}(\textbf{R}_{2},\textbf{R}_{1})=\frac{l_{0}}{3
\sqrt{2}}
\text{exp}(-l_{1}\textbf{R}_{1}^{2}-l_{2}\textbf{R}_{2}^{2}).\label{e3.42}
\end{eqnarray}
 Here
\begin{eqnarray}
l_{0}=(r+1)^{\frac{9}{4}}r^{\frac{-15}{8}}2^{\frac{3}{4}}(\pi \alpha
d^{2})^{\frac{-3}{2}}\label{e3.44}
\end{eqnarray}
\begin{eqnarray}
l_{1}=\frac{1}{4d^{2}}\Big(\frac{r+1}{2}\Big)^{2}\Big[\gamma-\frac{\beta^{2}}{\alpha}\Big]
\end{eqnarray}
\begin{eqnarray}
l_{2}=4\bar{k}+\frac{1}{2d^{2}}\sqrt{\frac{2r}{r+1}}\end{eqnarray}
where $\bar{k}=k_{f}b_{s}$,
\begin{eqnarray}
\alpha=8\bar{k}d^{2}\Big[\frac{r^{2}+1}{r^{2}}\Big]+1+r^{\frac{-3}{2}}\Bigg[\frac{(r+1)^{2}}{\sqrt{2(r+1)}}+1\Bigg]\end{eqnarray}
\begin{eqnarray}
\beta=8\bar{k}d^{2}\Big[\frac{r^{2}-1}{r^{2}}\Big]+1+r^{\frac{-3}{2}}\Bigg[\frac{r^{2}-1}{\sqrt{2(r+1)}}-1\Bigg]\end{eqnarray}
\begin{eqnarray}
\gamma=8\bar{k}d^{2}\Big[\frac{r^{2}+1}{r^{2}}\Big]+1+r^{\frac{-3}{2}}\Bigg[\frac{(r-1)^{2}}{\sqrt{2(r+1)}}+1\Bigg].\end{eqnarray}\label{e3.49}

Now for the off-diagonal kinetic energy kernel, eq. (\ref{e15})
gives \begin{equation} K_{kl}=\frac{1}{3}\sqrt{f} \Big(
\sum_{i=1}^{\overline{4}}-\frac{\underline{\nabla}_{i}^{2}}{2m_{i}}
\Big)\sqrt{f}.
\end{equation}
 Substituting in eq. (\ref{e3.14})
and using eq. (\ref{e3.26}) and eq. (\ref{e3.28}), we get

\begin{eqnarray}
\textbf{K}_{12}(\textbf{R}_{1},\textbf{R}_{2})=-\frac{l_{0}}{2m}\frac{1}{3
\sqrt{2}}\Big[r_{11}\textbf{R}_{1}^{2}+r_{12}\textbf{R}_{2}^{2}+r_{10}\Big]
\text{exp}(-l_{1}\textbf{R}_{1}^{2}-l_{2}\textbf{R}_{2}^{2})\label{e3.65a}
\end{eqnarray}
\begin{eqnarray}
\textbf{K}_{21}(\textbf{R}_{2},\textbf{R}_{1})=-\frac{l_{0}}{2m}\frac{1}{3
\sqrt{2}}\Big[r_{21}\textbf{R}_{1}^{2}+r_{22}\textbf{R}_{2}^{2}+r_{20}\Big]
\text{exp}(-l_{1}\textbf{R}_{1}^{2}-l_{2}\textbf{R}_{2}^{2})\label{e3.65b}
\end{eqnarray}
 where
\begin{eqnarray}
r_{11}=\Big(\frac{r+1}{2}\Big)^{4}\Bigg[\frac{8(r-1)^{2}}{(r+1)^{3}}\Bigg\{\Big(\frac{r-1}{r+1}\Big)
\Big(\frac{8\bar{k}}{(r-1)^2}+\frac{1+\sqrt{r}}{(r-1)^{2}d^{2}}\Big)-\Big(\frac{\beta}{\alpha}-\frac{r-1}{r+1}\Big)\nonumber\\
\Big(\frac{2\bar{k}}{r}+\frac{1}{2d^{2}\sqrt{r}(1+\sqrt{r})}\Big)\Bigg\}^{2}+\frac{32r}{(r+1)^{3}}\Bigg\{\Big(\frac{r-1}{r+1}\Big)
\Big(\frac{2\bar{k}}{r}+\frac{1}{2d^{2}\sqrt{r}(1+\sqrt{r})}\Big)\nonumber\\-\Big(\frac{\beta}{\alpha}-\frac{r-1}{r+1}\Big)
\Big(\bar{k}\frac{r^{2}+1}{r^{2}}+\frac{r^{-3/2}+1}{4d^{2}}\Big)\Bigg\}^{2}\Bigg]~~~~~~~~~~~~~~~~~~~~~~~~~~~~~~~~~~~~~~~~\label{e3.66}
\end{eqnarray}
\begin{eqnarray}
r_{10}=-\frac{3}{2} \left(\frac{r+1}{2}\right)^{2}
\Bigg\{8\frac{(r-1)^{2}}{(r+1)^{3}}
\left(\frac{8\bar{k}}{(r-1)^{2}}+\frac{1+\sqrt{r}}{(r-1)^{2}d^{2}}\right)+\frac{32
r}{(r+1)^{3}}~~~~~~~~~~~~~\nonumber\\ \left(\frac{\bar{k}
(r^{2}+1)}{r^{2}}+\frac{r^{\frac{-3}{2}}+1}{4
d^{2}}\right)\Bigg\}+\frac{3}{2} \frac{d^{2}}{\alpha}
(r+1)^{2}\Bigg\{\frac{8(r-1)^{2}}{(r+1)^{3} } \left(\frac{2
\bar{k}}{r}+\frac{1}{2 d^{2} \sqrt{r}
\left(1+\sqrt{r}\right)}\right)^{2}\nonumber\\+\frac{32
r}{(r+1)^{3}} \left(\frac{\bar{k}
(r^{2}+1)}{r^{2}}+\frac{r^{\frac{-3}{2}}+1}{4
d^{2}}\right)^{2}\Bigg\}-\frac{6(r+1)}{2 r} \left(2
\bar{k}+\frac{1}{2 d^{2}} \sqrt{\frac{2
r}{r+1}}\right)~~~~~\label{e3.67}
\end{eqnarray}
\begin{eqnarray}r_{12}=4 \left(\frac{r+1}{2r}\right)\left(2 \bar{k}+\frac{1}{2 d^{2}} \sqrt{\frac{2
r}{r+1}}\right)^{2}\label{e3.68}
\end{eqnarray}
\begin{eqnarray}r_{22}=r_{12}\label{e3.69}
\end{eqnarray}
\begin{eqnarray}r_{20}=\frac{8}{(r+1)^{2}} \Big(\frac{r+1}{2}\Big)^{2}\frac{24
d^{2}}{\alpha}\Bigg\{\Big(\bar{k}+\frac{1}{4d^{2}}\Big)^{2}+r
\Big(\frac{\bar{k}}{r^{2}}+\frac{r^{-\frac{3}{2}}}{4
d^{2}}\Big)^{2}\Bigg\}-6\frac{8}{(r+1)^{2}}\nonumber\\
\Big(\frac{r+1}{2}\Big)^{2}
\Bigg\{\Big(\bar{k}+\frac{1}{4d^{2}}\Big)+r
\Big(\frac{\bar{k}}{r^{2}}+\frac{r^{-\frac{3}{2}}}{4
d^{2}}\Big)\Bigg\}-6\frac{r+1}{2r}\Bigg\{2\bar{k}+\frac{1}{2d^{2}}
\sqrt{\frac{2r}{r+1}}\Bigg\}\label{e3.70}
\end{eqnarray}
 and
\begin{eqnarray}r_{21} = 2(r +
1)^2\Bigg\{\Big(1-\frac{\beta}{\alpha}\Big)^2 \Big(\bar{k} +
\frac{1}{4d^2}\Big)^2+
r\Big(1+\frac{\beta}{\alpha}\Big)^2\Big(\frac{\bar{k}}{r^2} +
\frac{r^\frac{-3}{2}}{4d^2}\Big)^2\Bigg\}.\label{e3.71}\end{eqnarray}
 Lastly for the potential energy
kernel with $k\neq l$, using eqs. \eqref{e16} and \eqref{e3.7b} in
eq. (\ref{e3.15}), changing variables and doing all the
integrations, we get
\begin{eqnarray}
\textbf{V}_{12}(\textbf{R}_{1},\textbf{R}_{2})=\textbf{V}_{21}(\textbf{R}_{2},\textbf{R}_{1})=l_{0}\Big[n_{1}\textbf{R}_{1}^{2}+n_{0}\Big]
\text{exp}(-l_{1}\textbf{R}_{1}^{2}-l_{2}\textbf{R}_{2}^{2}),\label{e3.75}
\end{eqnarray}
 with
\begin{eqnarray}
n_{0}=-\frac{8}{3}C\Big(\frac{r+1}{r}\Big)^{2}\frac{d^{2}}{\alpha}\label{e3.76}
\end{eqnarray}
\begin{eqnarray}
n_{1}=-\frac{4}{9}C\Big\{\frac{(r+1)^{4}}{4r^{2}}\Big\}\Bigg(\frac{\beta}{\alpha}-\frac{r-1}{r+1}\Bigg)^{2}.\label{e3.77}
\end{eqnarray}
 Putting expressions from eqs.
(\ref{e3.21}), (\ref{e3.24}), (\ref{Nkk}), (\ref{e3.65a}),
(\ref{e3.75}) and (\ref{e3.42}) in eq. (\ref{e3.13}), we get first
integral equation for $k$=1 and by putting expressions from eqs.
(\ref{e3.21}), (\ref{e3.24}), (\ref{Nkk}), (\ref{e3.65b}),
(\ref{e3.75}) and (\ref{e3.42}) in (\ref{e3.13}), we get the second
integral equation for $k=$2 that we have shown as  eq.
(\ref{e3.80}).

\section{}\label{app d}
Because of the spherical symmetry of the S-wave ($l=0$),
$\textbf{P}_{i}$ is replaced with $p_{i}$ (magnitude) with $i=1,2$.
Using the Parseval relation  eqs. (\ref{e3.91}) and (\ref{e3.92})
give
\begin{eqnarray}
A_{k}(u)=4\pi l_{0}\int
dp_{k}p_{k}^{2}F_{a}(p_{k},u)\chi_{k}(p_{k})\label{e3.103}\end{eqnarray}
\begin{eqnarray}
B_{k}(u)=4\pi l_{0}\int
dp_{k}p_{k}^{2}F_{b}(p_{k},u)\chi_{k}(p_{k}).\label{e3.104}\end{eqnarray}

Multiplying eq. (\ref{e3.125}) by $4\pi p_{1}^{2}F_{a}(p_{1},l_{1})$
and integrating w.r.t. $p_{1}$ and using eq. (\ref{e3.103}) we get
\begin{eqnarray}
\frac{A_{1}(l_{1})}{l_{0}}=4\pi
F_{a}(p_{c}(1),l_{1})-A_{2}(l_{2})W_{11}^{(1)}-B_{2}(l_{2})W_{12}^{(1)}\label{e3.105}\end{eqnarray}
Similarly multiplying eq. (\ref{e3.125}) by $4\pi
p_{1}^{2}F_{b}(p_{1},l_{1})$ and integrating w.r.t. $p_{1}$ and
using eq. (\ref{e3.104}), we get
\begin{eqnarray}
\frac{B_{1}(l_{1})}{l_{0}}=4\pi
F_{b}(p_{c}(1),l_{1})-A_{2}(l_{2})W_{21}^{(1)}-B_{2}(l_{2})W_{22}^{(1)}\label{e3.110}\end{eqnarray}
In the same way multiplying eq. (\ref{e3.126}) by $4\pi
p_{2}^{2}F_{a}(p_{2},l_{2})$ and $4\pi p_{2}^{2}F_{b}(p_{2},l_{2})$
and integrating w.r.t. $p_{2}$ and using eqs. (\ref{e3.103}) and
(\ref{e3.104}), we get
\begin{eqnarray}
\frac{A_{2}(l_{2})}{l_{0}}=-A_{1}(l_{1})W_{11}^{(2)}-B_{1}(l_{1})W_{12}^{(2)}\label{e3.114}\end{eqnarray}
\begin{eqnarray}
\frac{B_{2}(l_{2})}{l_{0}}=-A_{1}(l_{1})W_{21}^{(2)}-B_{1}(l_{1})W_{22}^{(2)}\label{e3.115}\end{eqnarray}
where W's in above equations depend upon $l^,s , n^,s$, $r^,s$,
${E}_c$, $\bar{C}$ and constituent quark mass of light quarks. Eqs.
\eqref{e3.105}, \eqref{e3.110}, \eqref{e3.114} and \eqref{e3.115}
can be written in the matrix form as follows
\begin{eqnarray}
WV_{1}=V_{2}\label{e3.120}
\end{eqnarray}
with
\begin{eqnarray}
W=\left(
  \begin{array}{cccc}
    l_{0}^{-1} & 0 & W_{11}^{(1)} & W_{12}^{(1)} \\
    0 & l_{0}^{-1} & W_{21}^{(1)} & W_{22}^{(1)} \\
    W_{11}^{(2)} & W_{12}^{(2)} & l_{0}^{-1} & 0 \\
    W_{21}^{(2)} & W_{22}^{(2)} & 0 & l_{0}^{-1} \\
  \end{array}
\right)\texttt{ }\label{e3.121}
\end{eqnarray}

\begin{eqnarray}
V_{1}=\left(
        \begin{array}{c}
            A_{1}(l_{1})\\
            B_{1}(l_{1})\\
            A_{2}(l_{2})\\
            B_{2}(l_{2})\\
        \end{array}
      \right)
\label{e3.122}
\end{eqnarray}
\begin{eqnarray}
V_{2}=4\pi\left(
        \begin{array}{c}
            F_{a}(p_{c}(1),l_{1})\\
            F_{b}(p_{c}(1),l_{1})\\
            0\\
            0\\
        \end{array}
      \right).
\label{e3.123}
\end{eqnarray}
From eq. \eqref{e3.120}), we can have
\begin{eqnarray}
V_{1}=W^{-1}V_{2} \label{e3.124}
\end{eqnarray}
which gives values of $A_{1}(l_{1}), B_{1}(l_{1}), A_{2}(l_{2})$ and
$B_{2}(l_{2})$ needed in eqs. (\ref{e3.132}) and (\ref{e3.133}).

\end{appendix}

\end{document}